\documentclass{article}
\usepackage{graphicx} 

\usepackage{jheppub} 
                     
\usepackage{booktabs,makecell, bm}
\usepackage[font=small]{caption}
\usepackage{framed}
\usepackage{hyperref}
\usepackage[export]{adjustbox}
\usepackage{amsmath}

\usepackage{url}
\usepackage[numbers]{natbib}
\DeclareMathOperator{\Tr}{Tr}
\usepackage{physics}
\usepackage{subcaption}

\title{Extremal AdS Black Holes as Fluids: \\A Matrix Large-Charge EFT Approach}

\author{Eunwoo Lee}

\affiliation{Department of Theoretical Physics, \\ Tata Institute
	of Fundamental Research, Homi Bhabha Rd, Mumbai 400005, India}
\emailAdd{eunwoo.lee@tifr.res.in}

\abstract{
We develop a simple, yet powerful, matrix-valued large‑charge EFT that captures the thermodynamic behavior of rotating extremal large-charge AdS black holes. 
We introduce a minimal “matrix EFT” by promoting the complex scalar in large charge EFT to a complex $N\times N$ adjoint scalar, whose $\mathcal{O}(N^2)$ modes contribute at zero temperature.  Employing a mean‑field approximation, we solve the self‑consistency equations and obtain explicit rigidly rotating fluid solutions.  We demonstrate that their energy, angular momenta, and charge densities exactly reproduce the thermodynamics and boundary stress tensor of zero‑temperature conformal fluids.  A microscopic mode‑counting further accounts for the $\mathcal{O}(N^2)$ entropy.
Via the fluid/gravity correspondence, this fluid describes an extremal AdS black hole in large charge limit. 
We also comment on supersymmetric BPS black holes, which fall outside the usual hydrodynamic regime but nevertheless exhibit simple, universal behavior in the large angular momentum limit. In this regime, their non-linear charge–spin relations simplify to form reminiscent of our extremal fluid solutions at large angular momentum limit.
}

\makeatletter
\gdef\@fpheader{}
\makeatother

\begin{document}

\maketitle

\section{Introduction}

Large black holes in AdS exhibit universal thermodynamic behavior. This universality can be effectively described by conformal fluid dynamics on the boundary of AdS spacetime (see, e.g., \cite{Policastro:2002se,Janik:2005zt,Son:2007vk,Bhattacharyya:2007vjd,Hubeny:2011hd} and references therein).
In particular, the authors of \cite{Bhattacharyya:2007vs} demonstrated that large rotating AdS black holes can be mapped to stationary solutions of relativistic conformal fluid dynamics on the boundary sphere. The stationary configuration corresponds to a rigidly rotating conformal fluid, and that both the thermodynamic properties and the boundary stress-energy tensor of the black holes are accurately captured by the behavior of this fluid solution.
Quite remarkably, this fluid/gravity correspondence holds not only for finite temperature black holes but also extends to certain extremal, non-supersymmetric configurations if a black hole carries large charge $Q$.

On the other hand, the large charge effective field theory (EFT) provides a systematic low-energy description of CFTs in sectors of fixed, large global charge $Q$ \cite{Hellerman:2015nra,Alvarez-Gaume:2016vff, Loukas:2016ckj, Loukas:2017lof, Hellerman:2017efx, Hellerman:2017veg, Hellerman:2017sur, Monin:2016jmo, Jafferis:2017zna, Bourget:2018obm, Hellerman:2018xpi, Cuomo:2019ejv,Badel:2019khk,Cuomo:2022kio,Cuomo:2023mxg,Badel:2022fya,Dondi:2024vua,Choi:2025tql,Watanabe:2025mnc}. In this framework, the Goldstone mode of the spontaneously broken global symmetry dominates the low-energy dynamics, and the effective action is organized as a controlled expansion in inverse powers of $Q$. The large charge EFT successfully captures the ground state structure, energy spectrum, and correlation functions within this fixed-charge sector.

The Wilsonian effective action describing the sector of a conformal field theory on $S^{d-1}\times \mathbb{R}$ with large fixed \( U(1) \) charge \( Q \) can be written, to leading order, as
\begin{align}\label{1}
\mathcal{L}=c|\partial\chi|^{d}+\text{subleading},
\end{align}
where \( \chi \) is the Goldstone mode associated with the spontaneously broken global \( U(1) \) symmetry, and \( c \) is a dimensionless constant.
Alternatively, this EFT can be equivalently formulated as a linear sigma model with the following Wilsonian action:
\begin{align}
    \mathcal{L}=-\partial^{\mu}\varphi\partial_{\mu}\varphi^*-m_d^2 \varphi\varphi^*-\lambda |\varphi|^{2d/(d-2)}+\text{subleading},
\end{align}
where \( \varphi \) is a complex scalar field charged under the global \( U(1) \) symmetry and $m_d$ is the conformal mass in $d$ dimension.
At leading order, both descriptions are equivalent. This can be seen by parameterizing \( \varphi = v\, e^{i\chi} \) and integrating out the radial mode \( v \). Substituting this back into the action yields the nonlinear Goldstone action \eqref{1}, up to corrections suppressed in the large-charge expansion \cite{Hellerman:2015nra}.

Recent studies have shown that the ground state of the EFT \eqref{1} at large charge $Q$ and large angular momentum $J$, in the regime \( 1 \ll Q \ll J \ll Q^{(d-1)/(d-2)}\), is well-approximated by a rigidly rotating fluid configuration~\cite{Cuomo:2017vzg,Cuomo:2019ejv,Choi:2025tql}. In this limit, the stress-energy tensor takes the same functional form as that of a zero-temperature conformal fluid undergoing rigid rotation \cite{Choi:2025tql}.
\footnote{We note that the stationary fluid description for extremal black holes is also valid only in $1 \ll Q \ll J \ll Q^{(d-1)/(d-2)}$. For $J\lesssim Q$, rigid rotation is not the relevant configuration. For instance, when \( J \lesssim Q^{1/(d-1)} \), the dynamics is dominated by low-energy phonon excitations on top of a homogeneous superfluid background. Also, the fluid solution cannot reach $J\gtrsim Q^{(d-1)/(d-2)}$. We elaborate on this correspondence in Section \ref{rev}.}

This fluid-like behavior in the large-charge sector naturally raises several questions.
\begin{itemize}
    \item Does there exist a universal hydrodynamic description applicable to the large-charge regime of conformal field theories?
    \item Is there a deeper correspondence between the large-charge ground state and the fluid dynamics of large-charge AdS extremal black holes?
    \item In particular, can extremal AdS black holes be understood as rigidly rotating, zero-temperature fluids that are dual to the ground states of large-charge effective field theories?
\end{itemize}
In a related context, the authors of \cite{Loukas:2018zjh} proposed a tentative correspondence, which they referred to as the AdS/EFT correspondence. They suggested that the ground state of the large-charge EFT at large \( Q \) may provide an effective description of extremal Reissner–Nordström black holes in AdS.

Nevertheless, the correspondence has not been understood completely because it is believed that the superfluid ground state described by the large charge EFT seems to be nearly unique, representing a low-entropy configuration dominated by a single macroscopic condensate. 
In contrast, the extremal AdS black hole, even at zero temperature, possesses a macroscopic degeneracy with an entropy that scales as $\mathcal{O}(N^2)$. 
Namely, the EFT action itself does not encode large numbers of microscopic degrees of freedom necessary to account for the vast entropy of the black hole. The large charge EFT does provide a description of the energy and stress tensor but does not capture the full microstate structure responsible for the black hole's entropy.

In this paper, we propose that extremal AdS black holes can be understood as large-charge ground states of a ``matrix" version of the large charge EFT, where \(\mathcal{O}(N^2)\) number of nearly degenerate modes are condensed.\footnote{Here, we are considering the classically extremal black hole. 
We are not taking into account quantum effects such as those governed by the Schwarzian mode \cite{Iliesiu:2020qvm}. Therefore entropy remains finite and nonvanishing at the zero temperature. We comment on this briefly in \ref{ent}.}
As a concrete example, we introduce a simple—but sufficiently general—extension of the large‑charge EFT by promoting the original \(U(1)\)‑charged complex scalar to a complex \(N\times N\) adjoint matrix.  In \(d=4\), our model is governed by the Lagrangian
\begin{align}\label{matrix lag}
 \mathcal{L}
 \;=\;
 -\Tr\Bigl(\partial^{\mu}X\,\partial_{\mu}\bar{X} + m_4^2\,X\bar{X}\Bigr)
 - V_4(X,\bar{X})\,,
\end{align}
where $X(\bar{X})=\Phi_1\pm i\Phi_2$, with $\Phi_1$ and $\Phi_2$ being $N\times N$ Hermitian matrices. A potential \(V_4(X,\bar X)\) is chosen to be quartic in $X,\bar{X}$, such as $
V_4 \sim \lambda\,\Tr\bigl(X\bar X X\bar X\bigr)
~\text{or}~
\lambda\,\bigl[\Tr(X\bar X)\bigr]^2,$ so as to preserve classical conformal invariance and the global \(U(1)\times SU(N)\) symmetry.  Although \(SU(N)\) appears here as a global symmetry, one may weakly gauge it to project onto \(SU(N)\) singlet states without affecting the large‑charge, large‑spin dynamics.  We return to this point in Section \ref{large}.

In this paper, we analyze \eqref{matrix lag} whose equation of motion is
$\!(-\partial_t^2+\nabla^2)X=\frac{\partial V}{\partial\bar{X}}$.
We employ a self‑consistent mean-field approximation,
$\frac{\partial V}{\partial\bar{X}}\approx CX$, reducing the equation of motion to the linear form
$(-\partial_t^2+\nabla^2)X=CX$. If we impose \( J_1 = J_2 \) for simplicity, the \( SO(4) \) symmetry ensures that \( C \) is constant on \( S^3 \). The equation of motion can then be solved using a mode expansion:
\begin{align}
X(t,\Omega)^{ij}
\approx\sum_{n,\ell}\frac{1}{\sqrt{2E_n}}\Bigl[a_{n\ell}^{ij}\,Y_{n\ell}(\Omega)\,e^{-iE_nt}
\;+\;b^{\dagger ij}_{n\ell}\,Y^*_{n\ell}(\Omega)\,e^{+iE_nt}\Bigr],
\end{align}
where $\;E_n=\sqrt{(n+1)^2+C}$, \(i,j=1,\dots,N\) label the matrix indices of \(X\), and $Y_{n\ell}$ the scalar harmonics of spin $\tfrac n2\otimes\tfrac n2$ on $S^3$.
Enforcing the self‐consistency condition then yields \(C\simeq\frac{\mu^2}{1-\Omega^2}\), and one finds that the solution predominantly occupies modes with
$n\approx \frac{\mu\Omega}{1-\Omega^2}$.

We compute the stress energy tensor, charge densities of this solution and confirm that they match the universal behavior of a rigidly rotating, zero-temperature conformal fluid. 
The solution also reproduces an entropy of order \(\mathcal{O}(N^2)\).

Although our model is a simple example, this adjoint-matrix setup captures key universal features of fluid dynamics exhibited by extremal black holes in AdS$_5$. Indeed, all known AdS$_5$/CFT$_4$ dual pairs involve adjoint matter fields, and we believe that this adjoint-matrix extension of large-charge EFT provides a useful and tractable framework for understanding the emergent fluid behavior of extremal AdS black holes from a microscopic, field-theoretic perspective.

As an aside, we comment on supersymmetric black holes in Section \ref{BPS}.  Although these solutions lie beyond the standard fluid/gravity regime, they exhibit simple behavior in the large–angular–momentum limit. Besides satisfying the linear BPS bound
\(E = \sum_i Q_i + \sum_a J_a\)
the black holes also obey a non‑trivial, non‑linear charge–spin constraint.  We find that in large angular momentum limit $J_a \gg Q_i$, the non-linear relation simplifies, as charge and spin follow a universal functional form reminiscent of that in fluid dynamics. This suggests that aspects of the universal large-spin behavior may persist even beyond the conventional hydrodynamic regime.

The paper is organized as follows:
In Section~\ref{rev}, we review the necessary background. 
In Section~\ref{fluid rev} we summarize the fluid/gravity correspondence for large rotating AdS black holes, and in Section~\ref{eft rev} we review the large‑charge EFT in fluid regime.
In Section~\ref{large}, we introduce our matrix generalization of the large‑charge EFT, solve the mean‑field equations, and derive explicit zero‑temperature fluid solutions that match extremal AdS\(_5\) black hole thermodynamics.
In Section~\ref{BPS}, we comment on the large‑spin limit of BPS black holes and the simplification of their non‑linear charge–spin relations.
Finally, in Section~\ref{dis}, we discuss extensions and open questions.



\section{Review}\label{rev}

In this section, we briefly review the essential background and derive some simple results
needed to motivate and contextualize our main results, about fluid description of large AdS black holes and the large-charge EFT framework.

\subsection{Large black holes as fluid}\label{fluid rev}
We begin by summarizing how large AdS black holes, particularly in the rotating case, can be effectively described by relativistic conformal fluids on the boundary. This correspondence is especially well understood in the hydrodynamic regime, where the derivative expansion is controlled and the stationary black hole geometry maps onto a rigidly rotating fluid on the boundary sphere.

A relativistic conformal fluid on $S^{d-1}\times \mathbb{R}$ carries energy, angular momenta and charges
\begin{align}
E,\quad J_a\;(a=1\ldots n=\lfloor d/2\rfloor),\quad Q_i\;(i=1\ldots c),
\end{align}
with conservation
\begin{align}
\nabla_\mu T^{\mu\nu}=0,\qquad \nabla_\mu J_i^\mu=0.
\end{align}
The conformal equation of state is given by
\begin{align}
\frac{1}{V}\ln Z_{\rm gc} \;=\; T^{d-1}\,h\!\bigl(\mu/T\bigr)\,,
\end{align}
where $\mu_i$ are chemical potentials and $h$ is fixed by static fluid (or static AdS black hole) data.

In \cite{Bhattacharyya:2007vs}, the authors show that stationary Navier–Stokes solutions on $S^{d-1} \times \mathbb{R}$ are given as rigidly rotating fluids on the sphere. It admits the velocity profile $u^{\mu}$ as
\begin{align}
u^t = \gamma, \quad u^{\varphi_a} = \gamma\, \Omega_a, \quad
\gamma \equiv \left(1 - \sum_a \Omega_a^2 \sin^2\theta_a\right)^{-1/2},
\end{align}
where \( \Omega_a \) are the angular velocities along orthogonal rotation planes labeled by \( \varphi_a \), and \( \theta_a \) are the polar angles associated with each plane in the embedding of \( S^{d-1} \). The factor \( \gamma \) ensures proper normalization of the relativistic velocity, \( u^\mu u_\mu = -1 \).
In equilibrium, all dissipative terms vanish, and the dynamics reduce to thermodynamic relations involving only \( T \), \( \mu_i \), and \( \Omega_a \).

The fluid approximation is valid when the mean free path
\begin{align}
\ell_{\rm mfp}\sim\frac{\eta}{\epsilon}
\end{align}
is much smaller than the curvature scale of the background—namely, the radius of \(S^{d-1}\). Here \(\eta\) denotes the shear viscosity and \(\epsilon\) the energy density of the fluid. 
In holographic theories, one can use the universal result \cite{Kovtun:2004de}
\begin{align}
\frac{\eta}{s} = \frac{1}{4\pi}.
\end{align}
Thus, requiring \(\ell_{\rm mfp}\ll1\) translates into the requirement that the dual AdS$_{d+1}$ black hole has a large horizon radius \( r_+ \gg \ell_{\text{AdS}} \). This follows from the scaling of energy and entropy in the gravitational description: \( E \sim r_+^d \), \( S \sim r_+^{d-1} \), implying
\begin{align}
\ell_{\rm mfp} \sim \frac{S}{E}\big|_{\Omega_a=0} \sim \frac{1}{r_+} \ll 1.
\end{align}
Hence, large AdS black holes correspond to fluids on \( S^{d-1} \times \mathbb{R} \) admitting a derivative expansion~\cite{Bhattacharyya:2007vs}.

The grand canonical partition function for the rotating fluid can be expressed by modifying the non-rotating fluid's partition function with a universal factor that depends on the angular velocities \cite{Bhattacharyya:2007vs,Banerjee:2012iz}:
\begin{align}\label{partition_fluid 2}
\log Z_{rot} = \log \left[\text{Tr} \exp \left(\frac{1}{T}(-H + \Omega_a J_a + \mu_i Q_i)\right)\right] = \frac{\log Z_{non-rot}}{\prod_{a=1}^{[d/2]}(1 - \Omega_a^2)}.
\end{align}

For concreteness, let us specialize to \( d = 4 \). The metric on the three-sphere \( S^3 \) is given by
\begin{align}
ds^2 = d\theta^2 + \sin^2\theta\, d\phi^2 + \cos^2\theta\, d\psi^2,
\end{align}
where \( \theta \in [0, \pi/2] \), and \( \phi, \psi \in [0, 2\pi) \) parameterize two orthogonal rotation planes.
A stationary, rigidly rotating fluid on \( S^3 \) has velocity field
\begin{align}
u^\mu = \gamma(1, 0, \Omega_1, \Omega_2), \quad 
\gamma = \left(1 - \Omega_1^2 \sin^2\theta - \Omega_2^2 \cos^2\theta\right)^{-1/2},
\end{align}
where \( \Omega_1 \) and \( \Omega_2 \) are the angular velocities along the \( \phi \) and \( \psi \) directions, respectively. 

The stress-energy tensor, charge current, and entropy current for a relativistic fluid are given by
\begin{equation}\label{eq: perfect fluid constituitive}
\begin{aligned}
T^{\mu\nu}_{\mathrm{perfect}} &= \epsilon\, u^\mu u^\nu + \mathcal{P}\, P^{\mu\nu}, \\
(J_i^{\mu})_{\mathrm{perfect}} &= \rho_i\, u^\mu, \\
(J_S^{\mu})_{\mathrm{perfect}} &= s\, u^\mu,
\end{aligned}
\end{equation}
where \( \epsilon \) is the energy density, \( \mathcal{P} \) is the pressure, \( \rho_i \) is the charge density associated with the \( i \)-th \( U(1) \) symmetry, and \( s \) is the entropy density. The projection tensor \( P^{\mu\nu} = g^{\mu\nu} + u^\mu u^\nu \) projects orthogonally to the fluid velocity. The subscript “perfect” indicates that these expressions correspond to the stationary (so that non-dissipative) fluid limit at leading order in the derivative expansion.

The charges for the fluid are given by integrating stress energy tensor and charge density functions over the sphere, and are written as follows.
\begin{equation}
\begin{aligned}\label{eq: charge relation}
    E&=\frac{V_3 T^4 h(\nu)}{(1-\Omega_1^2)(1-\Omega_2^2)}\left(\frac{2\Omega_1^2}{1-\Omega_1^2}+\frac{2\Omega_2^2}{1-\Omega_2^2}+3\right),\\
    S&=\frac{V_3 T^{3} [4h(\nu)-\nu_i h_i(\nu)]}{(1-\Omega_1^2)(1-\Omega_2^2)},\\
    J_1&=\frac{2V_3 T^4 h(\nu)\Omega_1}{(1-\Omega_1^2)^2(1-\Omega_2^2)},\\
    Q_i&=\frac{V_3 T^{3} h_i(\nu)}{(1-\Omega_1^2)(1-\Omega_2^2)}.
\end{aligned}
\end{equation}
where $V_3=Vol(S^3)=2\pi^2$, $\nu\equiv \frac{\mu}{T}$, and $h_i=\frac{\partial h(\nu)}{\partial \nu_i}$.
Note that \eqref{eq: charge relation} can also be derived from the partition function \eqref{partition_fluid 2} where $\log Z_{rot}=\frac{V_3 T^3 h(\nu)}{(1-\Omega_1^2)(1-\Omega_2^2)}$.

\subsubsection{Zero temperature limit}
For a charged rotating black holes, the large radius limit can be achieved even in the \emph{non‑supersymmetric extremal} limit ($T=0$, $\Omega_a,\mu_i$ finite). Fluid dynamics continues to match black hole data, while supersymmetric BPS solutions lie outside its regime of validity.
For the thermodynamic quantities in equation \eqref{eq: charge relation} to be well-defined in the \(T \rightarrow 0\) limit, the function \(h(\nu)\) should be expressed as follows:
\begin{align}
    16\pi G_5\cdot h(\nu) = \alpha_1\nu^4 + \alpha_2\nu^3 + \cdots = \alpha_1\left(\frac{\mu}{T}\right)^4 + \alpha_2\left(\frac{\mu}{T}\right)^3 + \cdots.
\end{align}
where $G_5$ is 5-dimensional Newton's constant. We assumed that there is only one internal charge $Q_1\equiv Q$, and one corresponding chemical potential $\mu$.
This allows us to write the partition function in terms of the chemical potential \(\mu\) and angular velocities \(\Omega_{1,2}\) as
\begin{equation}
    \log Z_{\text{rot}} = \frac{1}{T}\frac{V_3\left( \alpha_1 \mu^4 + T \alpha_2 \mu^3 + \cdots \right)}{16\pi G_5(1 - \Omega_1^2)(1 - \Omega_2^2)}.
\end{equation}
From this, the thermodynamic quantities follow as
\begin{equation}
\begin{aligned}\label{eq: charge relation 2}
    E &= \frac{V_3 \alpha_1 \mu^4}{16\pi G_5(1 - \Omega_1^2)(1 - \Omega_2^2)} \left( \frac{2\Omega_1^2}{1 - \Omega_1^2} + \frac{2\Omega_2^2}{1 - \Omega_2^2} + 3 \right), \\
    S &= \frac{V_3 \alpha_2 \mu^3}{16\pi G_5(1 - \Omega_1^2)(1 - \Omega_2^2)}, \\
    J_1 &= \frac{2V_3 \alpha_1 \mu^4 \Omega_1}{16\pi G_5(1 - \Omega_1^2)^2 (1 - \Omega_2^2)}, \\
    Q_i &= \frac{4V_3 \alpha_1 \mu^3}{16\pi G_5(1 - \Omega_1^2)(1 - \Omega_2^2)}.
\end{aligned}
\end{equation}
The relations in \eqref{eq: charge relation 2} hold in the hydrodynamic regime, where the mean free path satisfies $l_{\text{mfp}} \sim \frac{S}{E}|_{\Omega_a=0} \ll 1$, which is equivalent to $\mu \gg 1$. In addition, we assume the angular velocities are much larger than the inverse of the microscopic scale: $\Omega \gg \mathcal{O}(1/\mu)$. This ensures a coarse-grained fluid description. At smaller angular momenta, the appropriate description involves quantized phonon excitations rather than collective rigid rotation. The phonon excitations correspond to quasi-normal modes in the gravity dual.

In the small angular velocity limit, \(O(1/\mu)\ll\Omega_{1}, \Omega_{2} \ll 1\), the energy–charge–spin relation reduces to
\begin{align}\label{eq: sphere charge relation}
    E \approx c_1 Q^{4/3} + \frac{3}{4c_1} \frac{J_1^2 + J_2^2}{Q^{4/3}}+\text{subleading},
\end{align}
where $c_1 = \left(\frac{27}{512 \alpha \pi^2}\right)^{1/3}$. The ratio of the second to the first term on the RHS is
\[
\frac{J^2/Q^{4/3}}{Q^{4/3}} \sim  \mathcal{O}(\Omega^2)\ll 1.
\]
On the other hand the subleading term we neglect in \eqref{eq: sphere charge relation} scales as \(\sim Q^{2/3}\). This originates from the higher-derivative corrections from the gravity/fluid dynamics that we initially neglected \cite{Bhattacharyya:2007vs}. Therefore, the expansion \eqref{eq: sphere charge relation} is accurate only when \(J^2 \gg \mu^4 \sim Q^2\), i.e., in the regime \(J \gg Q\).

As angular momenta approach unity, the energy can be expressed in terms of other conserved charges as
\begin{align}\label{large charge exp}
        E \approx J_1 + J_2 + \frac{4c_1^3}{27} \frac{Q^4}{J_1 J_2}.
\end{align}  
Note that even as \(\Omega_{1,2} \to 1\), the angular momenta satisfy \(J \ll Q^{3/2}\). Therefore, in \(d = 4\), the fluid dynamics regime is valid in the window \(Q \ll J \ll Q^{3/2}\).

In Appendix~\ref{cclp}, we present a concrete example of a charged, rotating black hole in AdS$_5$~\cite{Chong:2005hr}, and demonstrate that extremal large black holes satisfy all the thermodynamic properties discussed above.

\subsection{Large‑Charge EFT \& Fluid Dynamics Regime}\label{eft rev}
In this subsection, we analyze the lowest energy state of a CFT\(_d\) on $S^{d-1}\times \mathbb{R}$ with a global $U(1)$ charge $Q$. The study focuses on the regime of large charge \(Q\gg1\) with angular momentum \(J\), employing a large charge EFT. 
A central idea of the large charge EFT is that fixing a large global charge \(Q\) in a CFT effectively introduces a new physical scale, even though the CFT is intrinsically scale-invariant \cite{Hellerman:2015nra}. This might seem paradoxical, but it arises because specifying a large charge \(Q\) on a compact space (like the sphere \(S^{d-1}\)) forces the system into a high-energy eigenstate. This state is not invariant under conformal transformations; it spontaneously breaks the conformal symmetry.

The validity of this EFT hinges on a clear separation of scales. As established in foundational works like \cite{Hellerman:2015nra}, the hierarchy is:
\begin{align}
\frac{1}{R} \ll \mu \ll \Lambda_{UV}
\end{align}
Here, \(1/R\) is the IR scale of the system set by the size of the sphere, \(\mu\) is the intermediate scale set by the charge density, and \(\Lambda_{UV}\) is the UV cutoff of the underlying CFT. 
The ground state for a fixed, large \(Q\) can be thought of as a dense medium of charge, akin to a superfluid. The density of charge, \(\rho \sim Q/R^{d-1}\), acts as an effective chemical potential, \(\mu \sim Q^{1/(d-1)}\). This chemical potential establishes a new energy scale, which allows for the construction of a low-energy EFT. 
The EFT expansion parameter is effectively the ratio of derivatives to the new scale, such as \(\partial/\mu\), which is small when the charge \(Q\) is large. This framework makes a perturbative, analytical study of a strongly-coupled CFT possible.

The EFT describes the dynamics of the Goldstone boson \(\chi\) associated with the spontaneously broken U(1) symmetry.
The leading-order Lagrangian for this EFT is given by:
\begin{align}\label{act}
    \mathcal{L}=c|\partial\chi|^{d}
\end{align}
where \(\chi\) is the Goldstone boson for the spontaneously broken U(1) symmetry and \(c\) is a model-dependent constant.
In the absence of angular momentum (\(J=0\)), the system settles into its simplest non-trivial ground state: a static, homogeneous superfluid completely covering the sphere \(S^{d-1}\).
In this state, the charge and energy are uniformly distributed. The superfluid is static, meaning its configuration doesn't change over time. In the EFT, this corresponds to a simple configuration for the Goldstone field: \(\chi(\mathbf{x}, t) = -\mu t\). The parameter \(\mu\) is the chemical potential, which sets the energy cost for adding a unit of charge.

By substituting this $\chi(\mathbf{x}, t) = -\mu t$ into the EFT Lagrangian \(\mathcal{L} = c|\partial\chi|^d\), one can derive the relationship between the total energy \(E\) (or scaling dimension \(\Delta = ER\)) and the total charge \(Q\). The calculation reveals a universal power-law dependence:
\begin{align}
    \Delta \approx c_1 Q^{d/(d-1)}
\end{align}
where \(c_1\) is a non-universal coefficient that depends on the specific CFT. The power $d/(d-1)$, however, is universal for all $d-$dimensioanl CFTs with a $U(1)$ symmetry.

The EFT can be improved by adding higher-derivative terms to \eqref{act}, which are suppressed by powers of the chemical potential \(\mu\). These terms generate a series of subleading corrections to the energy. The expansion takes the universal form: $\Delta = c_1 Q^{d/d-1} + c_2 Q^{d-2/d-1} + \cdots$.

The physical description of the the ground state for a given charge \(Q\) and angular momentum \(J\) undergoes several transitions as the relative scaling of \(J\) to \(Q\) changes.
\begin{itemize}
    \item In the regime $0 < J \ll Q^{1/(d-1)}$, the ground state is characterized by low-energy phonon excitations on top of the superfluid background. The energy receives a leading correction that scales with the angular momentum $J$, specifically $\delta \Delta \approx \sqrt{ \frac{J(J + d - 2)}{d - 1} }$, where $\delta \Delta$ represents the energy difference from the homogeneous ground state \cite{Hellerman:2015nra}.
    \item For $Q^{1/(d-1)} \ll J \lesssim Q$, the effective phonon description breaks down. Instead, the system minimizes its energy by nucleating codimension-two topological defects, analogous to vortices, which efficiently carry angular momentum. In the specific case of $d = 3$, this configuration corresponds to a vortex–anti-vortex pair situated on the two-sphere $S^2$ \cite{Cuomo:2017vzg}.
    \item For \( Q \ll J \ll Q^{(d-1)/(d-2)} \), the number of vortex-like defects increases and becomes densely populated \cite{Cuomo:2017vzg,Cuomo:2022kio,Choi:2025tql}. In a coarse-grained description, the ground state approximates a rigidly rotating fluid with angular velocity \( \Omega_a \). Importantly, this regime coincides with the fluid dynamics regime described in Subsection~\ref{fluid rev}. In particular, the stress-energy tensor and charge density take the same functional form as those of a zero-temperature, rigidly rotating conformal fluid \cite{Choi:2025tql}. Therefore, we refer to this intermediate regime as the \emph{fluid dynamics regime}.
    \item For \(Q^{(d-1)/(d-2)}\ll J\), the EFT breaks down and the system is expected to transition to the Regge limit, where the ground state is described by a collection of \(Q\) partons carrying the total angular momentum \(J\) \cite{Alday:2007mf,Komargodski:2012ek,Fitzpatrick:2012yx}.
\end{itemize}

\section{Large charge EFT for extremal black holes}\label{large}
In this section, we explore how large charge EFT captures the physics of extremal AdS black holes in the large-spin regime. We begin by analyzing the standard EFT and then develop its matrix-valued generalization, showing that it naturally leads to rigidly rotating fluid configurations at zero temperature.

\subsection{Large charge EFT and rigidly rotating fluid solutions}\label{fluid solution}

We begin by reviewing the standard large charge EFT, whose Wilsonian effective action describes the ground state at fixed large $U(1)$ charge. We focus on \( d = 4 \), where the leading-order effective Lagrangian takes the form
\begin{align}
    \mathcal{L} = \alpha |\partial \chi|^4.
\end{align}
\( \chi \) is the Goldstone mode associated with spontaneous breaking of the global \( U(1) \) symmetry.

The associated $U(1)$ current and stress-energy tensor are
\begin{align}\label{4dstress}
    j_{\mu} = 4\alpha |\partial\chi|^2 \partial_{\mu} \chi, \quad
    T_{\mu\nu} = 4\alpha |\partial\chi|^2 \partial_{\mu}\chi \partial_{\nu} \chi + g_{\mu\nu}\alpha |\partial\chi|^4.
\end{align}
The homogeneous ground state at large charge is described by \( \chi = -\mu t \), yielding
\begin{align}
    Q = 4\alpha \mu^3 \cdot V_3, \quad
    E = 3\alpha \mu^4 \cdot V_3,
\end{align}
and thus the scaling dimension
\begin{align}
    \Delta = ER = c_1 Q^{4/3}, 
\end{align}
where $c_1 = \left(\frac{27}{512 \alpha \pi^2}\right)^{1/3}.$
For small spin \( J \sim \mathcal{O}(1) \), low-energy excitations above the ground state are described by phonons. As \( J \sim \mathcal{O}(Q) \), it becomes favorable for the superfluid to form quantized vortex strings that carry angular momentum.

We are particularly interested in the ``fluid dynamics regime,'' defined by $Q \ll J_1, J_2 \ll Q^{3/2}$,
where vortex strings become densely packed. In this limit, the IR physics can be approximated by a rigidly rotating fluid as we explain below.

In the presence of vortex strings, the Goldstone field \( \chi \) becomes multi-valued. For a vortex string with winding number \( q \), the field satisfies
\begin{align}
    \oint d\chi = 2\pi q.
\end{align}
In the regime \( J \gg Q \), the vortex strings are sufficiently dense to be treated as a continuum. This coarse-grained description effectively allows \( q \) to take continuous values. In this picture \( \chi \) remains multi-valued, while its derivative \( \partial_\mu \chi \) is well-defined and smooth.
Therefore, instead of $\chi$, we treat \( \xi_\mu \equiv \partial_\mu \chi \) as independent fields~\cite{Choi:2025tql}. This relaxes the curl-free constraint on \( \xi \), allowing for a nonzero field strength \( F = d\xi \) that encodes the distribution of vortex defects.

The problem of determining the ground state thus becomes one of minimizing the energy with respect to \( \xi_\mu \), subject to fixed charge and angular momenta. Using the method of Lagrange multipliers, we extremize
\begin{align}\label{Lagrange multiplier}
    L \equiv E - \lambda_1 (J_1 - J^0_1) - \lambda_2 (J_2 - J^0_2) - \lambda_3 (Q - Q^0),
\end{align}
where the energy, charges, and angular momenta are computed via
\begin{align}
    E &= \int d\theta\, d\phi\, d\psi\, \sin\theta \cos\theta\, T_{tt}, \label{eq: E} \\
    J_1 &= \int d\theta\, d\phi\, d\psi\, \sin\theta \cos\theta\, (-T_{t\phi}), \quad
    J_2 = \int d\theta\, d\phi\, d\psi\, \sin\theta \cos\theta\, (-T_{t\psi}), \label{eq: J} \\
    Q &= \int d\theta\, d\phi\, d\psi\, \sin\theta \cos\theta\, j^t. \label{eq: Q}
\end{align}

Solving the variational equations yields:
\begin{align}\label{solsol}
    \xi^\mu = \mu\, \gamma^2 (1, 0, \Omega_1, \Omega_2), \quad
    \gamma = \left(1 - \Omega_1^2 \sin^2\theta - \Omega_2^2 \cos^2\theta \right)^{-1/2}.
\end{align}
The current and stress-energy tensor in terms of \( \xi^\mu \) now take the form
\begin{align}\label{st}
    j^\mu = 4\alpha |\xi|^2 \xi^\mu, \quad
    T_{\mu\nu} = \alpha\left(4|\xi|^2 \xi_\mu \xi_\nu + g_{\mu\nu} |\xi|^4 \right),
\end{align}
where \( |\xi| = \sqrt{-\xi^\mu \xi_\mu} \). 
This solution precisely matches that of a rigidly rotating relativistic fluid. The normalized fluid velocity is
\begin{align}
    u^\mu = \frac{\xi^\mu}{|\xi|} = \gamma (1, 0, \Omega_1, \Omega_2),
\end{align}
confirming the emergence of a perfect fluid behavior in this regime. The angular velocities \( \Omega_1, \Omega_2 \) describe rigid rotation on the sphere, with the Lorentz factor \( \gamma \) ensuring proper normalization.

The charges are given by integrating stress-energy tensor and charge densities given in \eqref{st}:
\begin{align}
    \frac{E}{V_3}&=\nonumber\frac{2(1-\Omega_1^2)+2(1-\Omega_2^2)-(1-\Omega_1^2)(1-\Omega_2^2)}{4(1-\Omega_1^2)^2(1-\Omega_2^2)^2}\alpha\mu^4\\
    \frac{J_1}{V_3}&=\nonumber\frac{\Omega_1 \alpha\mu^4}{2(1-\Omega_1^2)^2(1-\Omega_2^2)},\quad ~~
    \frac{J_2}{V_3}=\nonumber\frac{\Omega_2 \alpha\mu^4}{2(1-\Omega_1^2)(1-\Omega_2^2)^2}\\
    \frac{Q}{V_3}&=\frac{\alpha\mu^3}{2(1-\Omega_1^2)(1-\Omega_2^2)}
\end{align}
We use the following integral identity to obtain charges.
\begin{align}
    \int_0^{\pi/2}d\theta \sin\theta\cos\theta\frac{1}{(1-\Omega_1^2\sin^2\theta-\Omega_2^2\cos^2\theta)^3}=\frac{2-\Omega_1^2-\Omega_2^2}{4(1-\Omega_1^2)^2(1-\Omega_2^2)^2}
\end{align}
\begin{align}
    \int_0^{\pi/2}d\theta \sin\theta\cos\theta\frac{1}{(1-\Omega_1^2\sin^2\theta-\Omega_2^2\cos^2\theta)^2}=\frac{1}{2(1-\Omega_1^2)(1-\Omega_2^2)}
\end{align}
In the case of vanishing angular momenta \( J_1 = J_2 = 0 \), the energy of the ground state is given by
\begin{align}
     E\approx 2\pi^2\frac{3\mu^4}{2\lambda}=\frac{3}{2}\left(\frac{\lambda}{2\pi^2}\cdot \frac{1}{16}\right)^{1/3}\left(2\pi^2\frac{c^3}{\lambda}\right)^{4/3}=c_1Q^{4/3}
 \end{align}
In the regime of large angular momenta, \( J_1, J_2 \gg Q^{4/3} \), the energy takes the form
\begin{align}
     E-J_1-J_2\approx 2\pi^2\frac{\mu^4}{2\lambda(1-\Omega^2)^{2}}\approx\frac{\lambda}{4\pi^2}\cdot \frac{1}{16}\frac{Q^4}{J_1J_2}=\frac{4c_1^3}{27}\frac{Q^4}{J_1J_2},
\end{align}

\subsection{An equivalent sigma model description of large charge EFT}

Here, we consider an alternative but equivalent formulation of the large charge EFT in terms of a complex scalar with a relativistic sigma model–like action:
\begin{align}\label{sigact}
    \mathcal{L}_{\mathrm{IR}} = -\partial_{\mu} \varphi \partial^{\mu} \varphi^* 
    - m_4^2 \varphi \varphi^* 
    - \frac{\lambda}{2} (\varphi \varphi^*)^2,
\end{align}
where, in \( d = 4 \), the conformal mass is \( m_4 = 1 \), so the theory is classically conformally invariant.
The corresponding stress-energy tensor and charge density takes the form:
\begin{align}\label{stress 1}
    T_{\mu\nu}
    &= 2\partial_{(\mu} \varphi^* \partial_{\nu)} \varphi 
    - g_{\mu\nu} \left[
        \partial^{\rho} \varphi^* \partial_{\rho} \varphi 
        + \varphi^* \varphi 
        + \frac{\lambda}{2} (\varphi^* \varphi)^2
    \right] \nonumber\\
    &\quad + \frac{1}{3} \left( 
        R_{\mu\nu} + g_{\mu\nu} \nabla^2 
        - \nabla_\mu \nabla_\nu 
    \right) (\varphi^* \varphi),\\
    j_{\mu}&=\frac{1}{2i}\left(\varphi^{*}\partial_{\mu}\varphi-\varphi\partial_{\mu}\varphi^{*}\right)
\end{align}
where \( R_{\mu\nu} \) is the background curvature of the sphere.
The equation of motion is given as:
\begin{align}\label{eom}
    \frac{1}{\sqrt{-g}}\partial_{\mu}(\sqrt{-g}\partial^{\mu})\varphi&\approx\lambda (\varphi^*\varphi)\varphi
\end{align}

The equation of motion for the adjoint scalar field \( X_{ij} \) takes the form
\begin{align}\label{eom 3}
    \left(-\partial_t^2 + \nabla^2\right) \varphi = \lambda \, (\varphi^*\varphi) \, \varphi.
\end{align}
We now solve the equation of motion explicitly, rather than relying on an ansatz as in the previous section. \footnote{Of course, the equation of motion \eqref{eom 3} can be solved using an ansatz that is equivalent to \eqref{solsol}. This is shown in Appendix~\ref{equiv}.}
To proceed, we define a function \( C \) via the mean-field replacement
\begin{align}
    \lambda\langle \varphi^*\varphi \rangle \equiv C,
\end{align}
which reduces the equation of motion to a linear form:
\begin{align}
    \left(-\partial_t^2 + \nabla^2\right) \varphi = C \varphi.
\end{align}
This is a mean-field (or Hartree-type) approximation, and the resulting equation should be understood as a self-consistency condition: the field configuration used to compute \( \langle \varphi^*\varphi \rangle \) must solve this equation with the same value of \( C \).
\footnote{We believe approximation should be valid at large charge since other contributions are suppressed in powers of $1/Q\ll 1$.}

We consider $J_1=J_2$ for simplicity. It is natural to assume from the $SO(4)$ symmetry that $C$ is constant on the sphere.
The equation of motion can be solved exactly, and one can mode-expand the field operator as
\begin{align}\label{Xs}
    \varphi(t, \Omega) = \sum_{n,\ell} \frac{1}{\sqrt{2E_n}} \left[
a_{n\ell} Y_{n\ell}(\Omega) e^{-iE_n t}
+ b_{n\ell}^{\dagger} Y_{n\ell}^*(\Omega) e^{+iE_n t}
\right],
\end{align}
where $E_n = \sqrt{(n+1)^2+C}$, and $\Omega \in S^3$. 

The eigenfunctions of the Laplacian on $S^3$ are labeled by integers $n \in \mathbb{Z}_{\geq 0}$, and they transform under the $\mathrm{SU}(2)_L \times \mathrm{SU}(2)_R$ isometry group of the 3-sphere. The scalar harmonics are labeled by spins $(j_L, j_R)$ with
\begin{align}
j_L = j_R = \frac{n}{2}, \quad \text{and degeneracy } (n+1)^2.
\end{align}
where the degeneracy arises from the fact that \( m_{L,R} = -\frac{n}{2}, -\frac{n}{2}+1, \ldots, \frac{n}{2} \).
The Laplacian eigenvalue on $S^3$ is:
$$
-\nabla^2 Y_{n\ell} = \frac{n(n+2)}{R^2} Y_{n\ell}\;.
$$
Together with the conformal mass $1/R$, $(-\nabla^2+\frac{1}{R^2})Y_{n\ell}=\frac{(n+1)^2}{R^2}Y_{n\ell}$. 

Note that \( E_n \) denotes the energy of the \(n\)-th conformal descendant in the presence of the mean-field background. The corresponding anomalous dimension is given by \( E_n - (n+1) \), which is suppressed in the limit \( n \gg \sqrt{C} \). This suppression reflects the fact that for large angular momentum $n$, the effect of the mean-field potential (parametrized by \(C\)) becomes small and the spectrum becomes effectively free. 
Here, \( a_{n\ell} \) and \( b_{n\ell} \) are annihilation operators associated with the complex scalar field, labeled by quantum numbers \((n, \ell)\).

The occupation probability for a bosonic creation mode $a^{\dagger}$, labeled by quantum numbers $(n, \ell) = (n, m_1, m_2)$, is given by the Bose-Einstein distribution:
\begin{align}\label{bospart}
p_{n, m_1, m_2} = \frac{1}{e^{\beta (E_n - \Omega m_1 - \Omega m_2 - \mu)} - 1}.
\end{align}
Here, the magnetic quantum numbers $m_1 = m_L + m_R$ and $m_2 = m_L - m_R$ arise from the decomposition under the $\mathrm{SU}(2)_L \times \mathrm{SU}(2)_R$ isometry group of $S^3$.
We consider the zero-temperature limit $\beta \to \infty$, where excited-state contributions to the occupation number are exponentially suppressed. In this regime, only the modes satisfying
$E_n - \Omega m_1 - \Omega m_2 \approx \mu$
are macroscopically populated.
On the other hand, the conjugate modes created by $b^{\dagger}$ carry opposite $U(1)$ charge. Their occupation probability is given by
\begin{align}
\tilde{p}_{n, m_1, m_2} = \frac{1}{e^{\beta (E_n - \Omega m_1 - \Omega m_2 + \mu)} - 1},
\end{align}
which is always exponentially suppressed in the large-$\beta$ limit, as the exponent remains strictly positive. Thus, in the condensate regime, only the positively charged modes (created by $a^\dagger$) significantly contribute to the thermodynamics.

A crucial feature of the expression \eqref{bospart} is the possibility of divergence in the occupation number. Specifically, when the exponent in the denominator becomes small for certain modes, i.e.,
\begin{align}
    \beta (E_n -  \Omega m_1 -  \Omega m_2 -  \mu) \approx 0,
\end{align}
those modes will be highly populated. The occupation number for such modes can become large, scaling as
\begin{align}
    n_0 \sim \mathcal{O}\left( \frac{Q}{\# \text{ of condensate modes}} \right),
\end{align}
where \( Q \) is the total \( U(1) \) charge of the system. 
In this condensed regime, the dominant contribution to the partition function and thermodynamic quantities arises from these macroscopically occupied modes labeled by \( (n, m_1, m_2) \), while contributions from other states remain exponentially suppressed.
This is essentially the Bose-Einstein condensate.
 
Now, let us check whether a regime $\beta (E_n -  \Omega m_1 -  \Omega m_2 -  \mu) \approx 0$ exists.
This requires
\begin{align}
E_n = \sqrt{(n+1)^2 + C} \approx \Omega (m_1 + m_2) + \mu \leq \Omega n + \mu.
\end{align}
The following condition
\begin{align}\label{C}
C = \frac{(\mu - \Omega)^2}{1 - \Omega^2} \approx \frac{\mu^2}{1 - \Omega^2}
\end{align}
corresponds to the onset of the emergence of BEC modes. 

The equality selects modes with
\begin{align}
    n=\frac{\mu\Omega-1}{1-\Omega^2}
\end{align}
with $m_1+m_2=n.$
Therefore, the number of modes that condensate is given by $n\approx\frac{\mu\Omega}{1-\Omega^2}$.

To compute the mean-field expectation value \(\langle\varphi^*\varphi\rangle \), we use the mode expansion in equation~\eqref{Xs}. Each mode labeled by \((n, \ell)\) is orthogonal to the others and contributes with a weight of \(1/(2E_n)\), which can be approximated as \( (1 - \Omega^2)/(2\mu) \). 
If each mode is occupied by $n_0$ bosons, we estimate
\begin{align}
 \langle\varphi^*\varphi\rangle
&\approx \frac{n_0}{2E_n} \cdot \frac{ \mu \Omega}{1 - \Omega^2} = \frac{n_0\Omega}{2}.
\end{align}
The consistency condition then becomes
\begin{align}
C = \frac{n_0}{2} \approx \frac{\mu^2}{1 - \Omega^2},
\end{align}
which implies that each mode is occupied by 
$n_0 \approx \frac{2\mu^2}{\Omega(1 - \Omega^2)}.$

Substituting the mode expansion \eqref{Xs} into \eqref{stress 1} yields the following local densities:
\begin{align}\label{stress 2}
    T_{tt}&=\frac{\mu^4(3+\Omega^2)}{2\lambda(1-\Omega^2)^{3}}, \nonumber\\
    T_{t\phi}&=-\frac{2\mu^4\Omega\sin^2\theta}{\lambda(1-\Omega^2)^{3}},\nonumber\\
    T_{t\psi}&=-\frac{2\mu^4\Omega\cos^2\theta}{\lambda(1-\Omega^2)^{3}},\\
    \rho&=j^t=\frac{2\mu^3}{\lambda(1-\Omega^2)^{2}}\nonumber,
\end{align}
These expressions agree with those obtained at $\Omega_1 = \Omega_2 = \Omega$ in \eqref{st} using the ansatz, and also coincide with the corresponding results from fluid dynamics.\footnote{We use the following identity to obtain $T_{t\phi}$ and $T_{t\psi}$. 
\begin{align}\label{deriv1}
    \sum_{\substack{m_1+m_2=n,\\m_1, m_2\geq0}} m_1Y_{n,m_1,m_2}Y_{n,m_1,m_2}^{*}= n\sin^2\theta.
\end{align}
The derivation of \eqref{deriv1} is given in Appendix \ref{der}.
}
This is a constant angular velocity configuration because $j^\phi/j^t=\Omega$, $j^\psi/j^t=\Omega$. Namely, the angular velocities are $\Omega_1=\Omega_2=\Omega$. Of course, \eqref{stress 2} is consistent with the densities \eqref{st} obtained from the previous subsection.

For the case of unequal angular momenta, the same general procedure remains applicable. However, the analysis becomes technically more challenging, as the mean-field equation of motion is harder to solve when $\langle\varphi^*\varphi\rangle$ is no longer uniform over the sphere. We nevertheless expect the solution to still describe a rigidly rotating fluid configuration, as suggested by the ansatz-based results in Section~\ref{fluid solution} and Appendix~\ref{equiv}, though a detailed treatment is left for future work.

\subsection{Microscopic model for an extremal black hole}\label{micro}
Here, we consider a simple extension of the previous model by promoting the complex scalar field to a matrix-valued scalar field. Specifically, we take $X$ to be a complex scalar valued in the Lie algebra of $\mathfrak{su}(N)$. The Lagrangian for this model is given by
\begin{align}
 \mathcal{L}=-\text{Tr}\left(\partial^{\mu}X\partial_{\mu}\bar{X}+m_4^2 X\bar{X}\right)-\frac{\lambda}{2N^2}(\text{Tr}X\bar{X})^2
\end{align}
In principle, one could choose other forms of the potential $V(X, \bar{X})$ as long as they preserve the global $U(1)$ symmetry. For example, an alternative potential could be $V = \frac{\lambda}{N^2} \text{Tr}(X\bar{X}X\bar{X})$. However, the essential steps for analyzing the theory remain unchanged, and the approach outlined below can be applied to such variants as well.

The kinetic and mass terms in the Lagrangian resemble those appearing in $\mathcal{N}=4$ super Yang–Mills theory, where scalar fields transform under the adjoint representation of $SU(N)$ and carry $U(1)$ $R$-charges. In our setup, $SU(N)$ appears as a global symmetry rather than a gauge symmetry.
Nonetheless, we may \emph{weakly gauge} the $SU(N)$ symmetry to restrict our attention to singlet states. As we argue later in this subsection, this gauging does not significantly alter the physics of the model in the regime of interest.

The stress-energy tensor is written as
\begin{align}\label{eq:stress}
    T_{\mu\nu}
    &=2\Tr\partial_{\{\mu}\bar{X}\partial_{\nu\}}X
    -g_{\mu\nu}\bigg[
    \Tr\partial^{\mu}\bar{X}\partial_{\mu}X
    +V
    \bigg]
    \nonumber\\
    &\quad+\frac{1}{3}(R_{\mu\nu}
    +g_{\mu\nu}\nabla^2-\nabla_\mu\nabla_\nu)\textrm{Tr}(X\bar{X})
\end{align}
where $V=\Tr X\bar{X}+\frac{\lambda}{2N^2}(\Tr X\bar{X})^2$.

The equation of motion for the adjoint scalar field \( X_{ij} \) takes the form
\begin{align}
    \left(-\partial_t^2 + \nabla^2\right) X_{ij} = \frac{\lambda}{N^2} \, \text{Tr}(X \bar{X}) \, X_{ij}.
\end{align}
We define a function \( C \) via the mean-field replacement
\begin{align}
    \frac{\lambda}{N^2} \langle \text{Tr}(X \bar{X}) \rangle \equiv C,
\end{align}
which reduces the equation of motion to a linear form:
\begin{align}
    \left(-\partial_t^2 + \nabla^2\right) X_{ij} = C X_{ij}.
\end{align}

Since this matrix model can be viewed as essentially \( N^2 \) copies of the complex scalar theory discussed in the previous subsection,
the stress energy tensor and charge density are given as $N^2$ times the \eqref{stress 2}:
\begin{align}
    T_{tt}&=\frac{N^2\mu^4(3+\Omega^2)}{2\lambda(1-\Omega^2)^{3}}, \nonumber\\
    T_{t\phi}&=-\frac{2N^2\mu^4\Omega\sin^2\theta}{\lambda(1-\Omega^2)^{3}},\nonumber\\
    T_{t\psi}&=-\frac{2N^2\mu^4\Omega\cos^2\theta}{\lambda(1-\Omega^2)^{3}},\\
    \rho&=j^t=\frac{2N^2\mu^3}{\lambda(1-\Omega^2)^{2}}\nonumber,
\end{align}
The stress energy tensor and charge density coincides with the stress energy tensor and charge density of \eqref{eq: charge relation 2} if one sets $\frac{N^2}{\lambda}=\frac{\alpha_1}{16\pi G_5}$.

\subsubsection{Entropy}\label{ent}
One of the shortcomings of the mean‐field approximation is that we are studying an effective quadratic (Gaussian) theory. Such a “clean” spectrum has widely spaced levels due to the quantized single-particle spectrum $E_n$ and misses the enormous level density—and hence the entropy—generated by fluctuations around the saddle point and/or chaotic disorder in interaction.

Let us nevertheless naively treat our condensate modes as an ideal Bose gas in this quadratic theory. The entropy is given by
\begin{align}
S \;=\; \sum_{\substack{\text{modes}}}
\bigl[(n_0 + 1)\log(n_0 + 1)\;-\;n_0\log n_0\bigr]\approx N^2\frac{\mu\Omega}{1-\Omega^2}\log\left(\frac{2\mu^2}{\Omega(1-\Omega^2)}\right),
\end{align}
where the number of macroscopically occupied modes is
$\;N^2\,\tfrac{\mu\,\Omega}{1-\Omega^2}$\ and each has occupation
$\;n_0\approx\tfrac{2\mu^2}{\Omega(1-\Omega^2)}$.  
Imposing the $SU(N)$ singlet constraint—by weakly gauging the global symmetry—reduces the count of states by an $\mathcal O(N^2)$ factor (from the Haar‐measure). However the leading entropy still scales like
$ S\;\sim\;N^2\,\mu\,\ln\mu\;\gg\;N^2,$
so the singlet‐projection merely shifts the coefficient and does not alter the overall scaling.

Although the entropy scales as $\mathcal{O}(N^2)$, it remains parametrically smaller than the total charge \( Q \sim N^2 \mu^3 \). In contrast, the entropy of a large extremal AdS$_5$ black hole is of order \( Q \) itself (see equation~\eqref{eq: charge relation 2} and \eqref{cclp en}). Physically, this discrepancy arises because the Gaussian condensate in our mean-field description features macroscopically large occupation numbers \( n_0 \sim \mu^2 \gg 1 \) per mode. In a genuine black hole microstate ensemble, strong interactions lift these degeneracies and scramble the single-particle spectrum, yielding an entropy of order \( Q \), consistent with expectations from maximal chaos.

To understand how to recover black hole–like entropy from our model, we believe that one must go beyond the leading mean-field approximation and incorporate fluctuations and higher derivative terms that were previously neglected.
One of the possible ways to model the necessary level repulsion and chaotic dynamics is to introduce a random quartic interaction among the matrix modes. For instance, consider the Lagrangian
\begin{align}
 \mathcal{L} = -\Tr\left(\partial^\mu X\,\partial_\mu \bar{X} + m_4^2\,X\bar{X}\right)
   - \frac{1}{2N^2}\sum_{i,j,k,l}
     \lambda_{ijkl}\,X_{ij}\,\bar{X}_{ij}\,X_{kl}\,\bar{X}_{kl},
\end{align}
where the couplings $\lambda_{ijkl}$ are drawn from a Gaussian ensemble with mean $\lambda$ and variance scaled appropriately with \( N \) and/or \( \mu \). This construction is motivated by matrix generalizations of the SYK model~\cite{Sachdev:1992fk,Maldacena:2016hyu, cotler2017black}. We expect that such interactions would induce level repulsion and chaotic mixing, raising the entropy to \( \mathcal{O}(Q) \), as anticipated for a strongly coupled system with holographic dual.

Of course, at strictly zero temperature the entropy must eventually vanish due to quantum effects, such as the emergence of a Schwarzian mode (See, for example,~\cite{Iliesiu:2020qvm} for a detailed discussion.). It would be very interesting to compute the entropy in this extended model quantitatively and study how it interpolates from \( \mathcal{O}(N^2\mu^3) \) down to zero as \( T \to 0 \). We leave this investigation for future work.

\section{Comments on BPS Black Holes at Large Angular Momentum}\label{BPS}

In this section, we discuss the limiting behavior of BPS black holes when the angular momenta satisfy \(J_a \gg Q_i\).  Although the standard fluid/gravity correspondence does not directly apply to BPS solutions \cite{Bhattacharyya:2007vs}, one can nonetheless identify some universal features in the large‑spin regime.
Notably, BPS black holes obey both the linear BPS bound
\begin{align}
E = \sum_i Q_i + \sum_a J_a
\end{align}
and a non‑linear constraint among the charges and angular momenta.  We will see in the remainder of this section, that one can derive an approximate form of the non‑linear charge constraint in the large‑spin limit \(J_a \gg Q_i\).

It is well known that all BPS black holes satisfy $\mu = 1$ and $\Omega = 1$. The requirement $\Omega = 1$ follows from general principles of conformal field theory. Specifically, in a unitary CFT, angular chemical potentials $\Omega > 1$ lead to divergences in the partition function due to the unbounded contribution from angular descendants. These descendants always exist, and when $\Omega > 1$, their Boltzmann weights grow exponentially, rendering the partition function ill-defined.

On the other hand, any configuration with \(\Omega < 1\) cannot correspond to a BPS state.  In the strict zero‑temperature limit \(T \to 0\), all descendant states are exponentially suppressed.  To keep these descendants unsuppressed as \(T \to 0\), one must have \(\Omega = 1\), so that BPS descendants continue to contribute to the partition function.

With this in mind, let us approach the limit \(\Omega \to 1\).  In this regime, the equilibrium thermal partition function at finite temperature can be written as\footnote{See \cite{Banerjee:2012iz,Bajaj:2024utv} for discussions on the universal denominator structure of the thermal partition function at large angular momentum. Notably, this structure persists even outside the fluid dynamic regime. We thank Shiraz Minwalla for helpful discussions.}
\begin{align}\label{tpart}
    \ln Z \;=\; \ln\!\bigl(\Tr\,e^{-\beta(E - \Omega_a J_a - \mu_i Q_i)}\bigr)
    \;=\;
    T^{\,d-1}\,\frac{h(\mu_i, T)}{\displaystyle\prod_a\bigl(1 - \Omega_a^2\bigr)}
    \;+\;\text{subleading}\,,
\end{align}
where \(h(\mu_i, T)\) is a function of the chemical potentials and temperature.  The universal \(1/\prod_a(1-\Omega_a^2)\) dependence follows from the proliferation of conformal descendants as \(\Omega_a \to 1\).

We now consider the joint limit \( T \to 0 \) and \( \Omega \to 1 \), while keeping the scaling such that \( \beta(1-\Omega) \ll 1 \). In this regime, we assume that the leading behavior of the partition function at low temperature is given by
\begin{align}\label{eq:extremal well}
    \ln Z
    = \frac{1}{T}\frac{V_{d-1} \, g(\mu_i)}{\prod_a(1-\Omega_a^2)}
    \left(1 + \mathcal{O}(1 - \Omega_a, T)\right).
\end{align}
This scaling is assuming a nontrivial zero-temperature thermodynamics, as thermodynamic quantities such as energy, angular momenta, charge, and entropy remain finite in this limit.
For example in $d=4$, thermodynamic quantities in the small-temperature limit take the form:
\begin{align}\label{bps charge}
\begin{split}
    E&=\frac{V_3 g(\mu)}{(1-\Omega_1^2)(1-\Omega_2^2)}\left[2\frac{\Omega_1^2}{1-\Omega_1^2}+2\frac{\Omega_2^2}{1-\Omega_2^2}-1\right]+\mu_i Q_i,\\
    J_a&=\frac{V_3 g(\mu)}{(1-\Omega_1^2)(1-\Omega_2^2)}\left[\frac{2\Omega_a}{1-\Omega_a^2}\right],~~~(a=1,2),\\
    Q_i&=\frac{V_3 \partial_{\mu_i}g(\mu)}{(1-\Omega_1^2)(1-\Omega_2^2)}
\end{split}
\end{align}

In order to saturate the BPS bound $E=J_1+J_2+\sum_iQ_i$, the chemical potential dependence must satisfy
\begin{align}
    \sum_i(\mu_i-1)\partial_{\mu_i}g(\mu)=3g(\mu)+\mathcal{O}((1-\Omega_1),(1-\Omega_2)).
\end{align}
This implies that $g(\mu)$ must be a homogeneous function of degree $3$ in the shifted chemical potentials $\mu_i-1$.

Moreover, in the BPS sector, we require the thermodynamic quantities \(E\), \(J_a\), and \(Q_i\) to remain finite.  
To achieve this, the chemical potentials must approach unity, \( \mu_i \to 1 \),  
in order to counterbalance the divergence arising from the factor \(1 - \Omega_a^2\).

Therefore, the partition function at BPS limit at large angular momentum limit is written as
\begin{align}
    \ln Z \approx a\frac{g(\delta_i)}{\omega_1\omega_2}
\end{align}
where $a$ is a theory dependent constant. We define the normalized angular chemical potential 
\(\omega \equiv \beta(1 - \Omega)\) and normalized charge potential \(\delta \equiv \beta(1 - \mu)\). Therefore, $g(\delta_i)$ is a degree 3 homogeneous function in $\delta_i$. 

Since the numerator also vanishes in this limit, one might wonder whether the subleading terms neglected in~\eqref{tpart} could become relevant.  
For example, one could speculate whether the partition function admits an alternative scaling behavior arising from such subleading terms, such as  
\begin{align}
    \log Z \sim \frac{1}{T}\frac{(\mu - 1)^2}{1 - \Omega^2}.
\end{align}
However, this possibility is ruled out by the regime we are working in: a large angular momentum limit where the normalized angular chemical potential \(\omega\) is much smaller than the normalized charge potential \(\delta\).  
In other words, we are taking a scaling limit with \(\omega \ll \delta\), reflecting the dominance of rotational contributions over those from the charge sector.
\footnote{
In the BPS limit, the partition function can be written in terms of renormalized chemical potentials as
\begin{align}
    Z_{\text{BPS}} = \Tr_{\text{BPS}} \left[ e^{-\omega_a J_a - \delta_i Q_i} \right],
\end{align}
where the trace receives contributions only from operators saturating the BPS bound.}

For equal charge chemical potentials, angular momenta and charge is given by   
\begin{align}\label{BPS charges}
    J_1= a\frac{\delta^3}{\omega_1^2\omega_2},\quad
    J_2= a\frac{\delta^3}{\omega_1\omega_2^2},\quad
    Q\equiv \sum_iQ_i= a\frac{3\delta^2}{\omega_1\omega_2}
\end{align}
Therefore, we obtain the following non-linear charge relation at large angular momentum.
\begin{align}
    J_1J_2=\frac{1}{27a}Q^3.
\end{align}

Now we apply these results to $\mathcal{N}=4$ SYM, whose holographic dual is type IIB supergravity on AdS$_5\times S^5$.  In this setup one finds BPS black‐hole solutions \cite{Gutowski:2004ez,Chong:2004na,Chong:2005hr} that obey
$
E \;=\; Q_1 + Q_2 + Q_3 + J_1 + J_2,
$
but exist only on a codimension‑one hypersurface in the five‑dimensional charge space.  Equivalently, in terms of the intensive variables
$
q_i = \frac{Q_i}{N^2},~ j_a = \frac{J_a}{N^2},
$
the BPS charges satisfy the nonlinear constraint 
$
q_1q_2q_3 + \tfrac12\,j_1j_2
\;=\;
\bigl(q_1 + q_2 + q_3 + \tfrac12\bigr)\,
\bigl(q_1q_2 + q_2q_3 + q_3q_1 - \tfrac{j_1 + j_2}{2}\bigr)$ \cite{Kim:2006he}. \footnote{See also \cite{Larsen:2024fmp,Chang:2024lxt} for recent field‑theoretic arguments explaining the emergence of non‑linear charge–spin relations.}

In the large‑charge, large‑spin limit ($q_i,j_a\gg1$), this simplifies to
\begin{align}
\frac{j_1j_2}{2}\;\approx\;(q_1 + q_2)\,(q_2 + q_3)\,(q_3 + q_1).
\end{align}
It turns out that the following homogeneous function $g(\delta)$ reproduce the relation above.
\begin{align}
g(\delta)=
\frac{N^2}{16}(\delta_1+\delta_2-\delta_3)\,
(\delta_2+\delta_3-\delta_1)\,
(\delta_3+\delta_1-\delta_2)\,,
\end{align}
so that the free energy becomes
\begin{align}
\ln Z=\frac{N^2}{16}\frac{(\delta_1+\delta_2-\delta_3)\,(\delta_2+\delta_3-\delta_1)\,(\delta_3+\delta_1-\delta_2)}
{\omega_1\,\omega_2}\,.
\end{align}
Here recall that take $\omega_a$ small compared to $\delta_i$, corresponding to large angular momentum regime.

Finally, to extract the entropy one must include the first subleading term in the small‑$T$ expansion,
\begin{align}
\frac{1}{\beta}\ln Z
=\frac{V_{d-1}\bigl(g(\mu_i)+f(\mu_i)\,T\bigr)}{\prod_{a}(1-\Omega_a^2)}+\cdots,
\end{align}
so that at $T=0$
\begin{align}
S
=\frac{V_{d-1}\,f(\mu_i)}{\prod_{a}(1-\Omega_a^2)}.
\end{align}
To prevent $S$ from diverging as $\Omega_a\to1$, it is natural that $f(\mu_i)\sim(\mu_i-1)^2$ for $d=4$.  Hence the entropy scales like $S\propto N^2\frac{\delta^2}{\omega_1\omega_2}$ in the BPS limit.

It is also important to emphasize that the BPS black hole cannot be thought of as a large black hole with angular velocities approaching $\Omega_1, \Omega_2 \to 1$, since in the BPS case, the chemical potentials $\mu_i$ are fixed to $1$ and no longer behave as large parameters. As a result, the BPS regime lies outside the scope of the large-charge, large-angular-momentum expansion considered in Section \ref{large}.

Even if fluid dynamics does not strictly apply for BPS black holes, all thermodynamic quantities exhibit a universal functional dependence on \(\Omega\).  
This arises from the structure of the partition function, which takes the form \(\ln Z \propto \frac{1}{\prod_a (1 - \Omega_a^2)}\).  
Such behavior suggests that certain structural features of the charge–spin relations persist beyond the hydrodynamic regime,  
potentially pointing to a deeper universality that extends into the BPS limit.  
It would be interesting to investigate whether these features can be derived from an appropriately modified version of the supersymmetric large-charge EFT  \cite{Hellerman:2017sur,Cuomo:2024fuy}.

\section{Discussion}\label{dis}

In this work, we developed an effective field theory framework to describe the zero-temperature, large-charge sector of strongly interacting CFTs in four dimensions, based on a matrix-valued scalar field theory. Motivated by the thermodynamics of extremal AdS black holes—which carry large charge and angular momentum—we investigated whether such black holes can be viewed as the ground states of this EFT. 

To this end, we formulated a classically conformal, $U(1) \times SU(N)$-invariant matrix model with quartic interactions and analyzed its saddle point in the large-charge, large-spin regime. Using a mean-field approximation, we identified the dominant condensate modes as spherical harmonics on $S^3$ aligned with the rotation, carrying both $U(1)$ charge and angular momentum. These modes collectively realize a rigidly rotating fluid configuration. The associated stress-energy tensor and conserved charges precisely match those of extremal AdS black holes in the corresponding limit. This supports the interpretation of this EFT ground state as an extremal AdS black hole in the hydrodynamic limit.

However, the mean-field approximation underestimates the entropy of black holes: although the total entropy scales as $\mathcal{O}(N^2)$, it remains parametrically smaller than the charge $Q \sim N^2 \mu^3$. This discrepancy arises because each spherical harmonics mode has a large occupation number, leading to limited entropy per mode. To model the chaotic dynamics and level repulsion expected in the microscopic black hole spectrum, we proposed augmenting the theory with a disordered quartic interaction, inspired by matrix generalizations of the SYK model. We expect that such interactions will lift the degeneracy of the single-particle spectrum and enhance the entropy, potentially recovering the gravitational result $S \sim Q$. A quantitative analysis of this mechanism is left for future work.

Finally, we briefly commented on the BPS limit, in which supersymmetric black holes obey a nonlinear charge–spin relation. Although the fluid dynamics framework does not strictly apply in this regime, we observed that in the limit $\Omega \to 1$, the charge–spin relation obtained from our EFT resembles that of the BPS bound. This suggests a potentially universal structure governing large angular momentum states across different sectors, meriting further investigation.

Several avenues for future investigation present themselves:

\begin{itemize}
  \item \textbf{General dimensions and unequal spins.}  Our analysis was simplified by specializing to \(d=4\) and imposing \(\Omega_1=\Omega_2\).  Extending the mean‑field computation to \(d=3\) (where only a single angular momentum exists) or to \(d=4\) with \(\Omega_1\neq\Omega_2\) will be more involved, but we expect the solution would be described by the rigidly rotating fluid. It would be interesting to check whether this is true.
  
  \item \textbf{Beyond leading order.}  We have restricted ourselves to leading‑order (classical) thermodynamics.  Computing subleading \(1/Q\), \(1/N\), or derivative corrections in the EFT would improve the match to finite‑size and quantum corrections of near‑extremal black holes.
  
  \item \textbf{Enriched field content.}  Here we considered only a single adjoint scalar.  Incorporating additional fields—such as fermions \cite{Komargodski:2021zzy, Dondi:2022zna}, gauge fields, or multiple interacting matrices—would bring the model closer to realistic AdS/CFT examples (e.g.\ \(\mathcal{N}=4\) SYM).
  
  \item \textbf{Excitations and quasi‑normal modes.}  While we focused on the ground‑state (condensate) solution, studying linearized fluctuations around this background will reveal the spectrum of quasi‑normal modes in the dual black hole and shed light on relaxation and transport in the extremal limit.
  
  \item \textbf{Exact zero‑temperature and quantum effects.}  Our framework describes a near‑extremal regime with small but nonzero temperature.  At exactly \(T=0\), quantum (Schwarzian-like) effects become important and the entropy should vanish. 
  It would be valuable to extend the EFT to include these quantum corrections and reproduce the complete zero‑temperature behavior.
  
  \item \textbf{BPS black holes.}  We have briefly commented on BPS charge–spin relations in the large‑spin limit.  A natural next step is to construct a large‑charge matrix model whose ground state exactly captures the extremal BPS black hole spectrum, potentially illuminating microstate counting for supersymmetric AdS black holes.
\end{itemize}

Overall, the matrix large-charge EFT offers a tractable, field-theoretic framework for capturing the hydrodynamic behavior of extremal AdS black holes. We expect that the ideas developed here will find broader relevance across both holographic systems and strongly correlated phases in condensed matter physics.

\section*{Acknowledgments}
We would like to thank Shiraz Minwalla and Chintan Patel for valuable discussions and insightful comments. We also thank Jaehyeok Choi for collaboration during the early stages of this project.
The work of E.L. was supported by Basic Science Research Program through the National Research Foundation of Korea (NRF) funded by the Ministry of Education RS-2024-00405516 and the Infosys Endowment for the study of the Quantum Structure of Spacetime.

\appendix

\section{Example: Extremal AdS Black Hole as a Fluid}\label{cclp}

In this appendix, we present an explicit example of a charged rotating AdS$_5$ black hole solution that exhibits the thermodynamic properties predicted by \cite{Bhattacharyya:2007vs}. Specifically, we use the extremal limit of the black hole solution constructed in \cite{Chong:2005hr} as a benchmark.
The mass, angular momenta, and charge of the general solution are given by
\begin{align}\label{chong charge}
    E &= \frac{1}{G}\frac{\pi\left[ m(2\Xi_a + 2\Xi_b - \Xi_a\Xi_b) + 2q ab(\Xi_a + \Xi_b) \right]}{4\Xi_a^2\Xi_b^2}, \nonumber \\
    J_1 &= \frac{1}{G} \frac{\pi\left[2am + qb(1 + a^2) \right]}{4\Xi_a^2\Xi_b}, \nonumber \\
    Q &= \frac{1}{G} \frac{3\pi q}{4\Xi_a\Xi_b} = Q_1 + Q_2 + Q_3,
\end{align}
where \( \Xi_a = 1 - a^2 \), \( \Xi_b = 1 - b^2 \), and \(G\) is the five-dimensional Newton constant.

In the extremal limit, the parameters scale as
\begin{align}
    q^2 \approx 2r_+^6, \quad m \approx \frac{3r_+^4}{2},
\end{align}
where $r_+$ refers to the radius of the black hole horizon.
Substituting the parameters into the expression \eqref{chong charge}, we obtain the thermodynamic quantities near extremality:
\begin{align}\label{cclp en}
    E &\approx \frac{1}{G}\frac{3\pi r_+^4(2\Xi_a + 2\Xi_b - \Xi_a\Xi_b)}{8\Xi_a^2\Xi_b^2}, \nonumber \\
    J_1 &\approx \frac{1}{G}\frac{3\pi N^2 r_+^2 a}{4\Xi_a^2\Xi_b}, \nonumber \\
    Q &\approx \frac{1}{G} \frac{3\sqrt{2}\pi r_+^3}{4\Xi_a\Xi_b}, \nonumber \\
    S &\approx \frac{1}{G} \frac{\pi^2 r_+^3}{2\Xi_a\Xi_b} = \frac{\sqrt{2}\pi}{3} Q.
\end{align}

For the special case of zero angular momenta (\(J_1 = J_2 = 0\)), the energy and charge reduce to
\begin{align}
    E &= \frac{1}{G}\frac{3\pi m}{4} \approx \frac{9\pi r_+^4}{8G}, \nonumber \\
    Q &= \frac{1}{G} \frac{3\pi q}{4} \approx \frac{3\sqrt{2} \pi r_+^3}{4G}.
\end{align}
Solving for \(E(Q)\), we find the characteristic large-charge scaling:
\begin{align}
    E \approx \frac{9\pi}{8G} \left( \frac{4G}{3\sqrt{2}\pi} \right)^{4/3} Q^{4/3} \equiv c_1 Q^{4/3}.
\end{align}
In the fast-rotation limit \(a, b \rightarrow 1\), the energy behaves as
\begin{align}
    \delta E \equiv E - J_1 - J_2 \approx \frac{1}{G} \frac{3\pi r_+^4}{8} \frac{1}{\Xi_a\Xi_b} \approx \frac{G}{6\pi} \frac{Q^4}{J_1 J_2} = \frac{4c_1^3}{27} \frac{Q^4}{J_1 J_2}.
\end{align}
This scaling matches the expression for the energy obtained from \eqref{large charge exp}.

\section{Ansatz-Based Solution of the Sigma Model}\label{equiv}
In this Appendix, we analyze the equations of motion \eqref{eom} using the following ansatz, which will turn out to be equivalent to the vortex-fluid solution discussed earlier:
\begin{align}\label{ansatz}
    X = u \cdot \mathrm{Pexp} \left[
        i \int f \left(
            dt - \Omega_1 \sin^2\theta\, d\phi 
            - \Omega_2 \cos^2\theta\, d\psi 
        \right)
    \right],
\end{align}
with
\begin{align}\label{ans}
    f = \frac{\mu}{1 - \Omega_1^2 \sin^2\theta - \Omega_2^2 \cos^2\theta}.
\end{align}
Here, \( \mathrm{Pexp} \) denotes the path-ordered exponential.

At first glance, the ansatz \eqref{ansatz} may seem ill-defined, as it picks up a nontrivial phase around closed loops on the sphere. However, this is precisely the same phenomenon encountered in the effective vortex description, where the phase field \( \chi \) becomes multi-valued due to the presence of dense vortex defects, and only its derivative \( \xi_\mu = \partial_\mu \chi \) remains well-defined.
In other words, the non-single-valuedness of \( X \) reflects the same underlying physics: a coarse-grained description of a state with vorticity, now encoded in the angular dependence of \( f \).

The equation of motion is written as
\begin{align}
    f^2(1-\Omega_1^2\sin^2\theta-\Omega_2^2\cos^2\theta)\approx\lambda \varphi\varphi^*
\end{align}
Therefore, $u$ in \eqref{ansatz} is given by
\begin{align}\label{ans sol}
    u\approx \frac{\mu}{\sqrt{\lambda}\sqrt{1-\Omega_1^2\sin^2\theta-\Omega_2^2\cos^2\theta}}
\end{align}
Note that we neglected derivatives in $\theta$ in \eqref{eom} since the following terms
\begin{align}
    &\partial_{\theta}u \propto \frac{(a^2-b^2)\cos\theta\sin\theta}{(1-\Omega_1^2\sin^2\theta-\Omega_2^2\cos^2\theta)^{3/2}}\sim O\left(\frac{1}{(1-\Omega_1^2\sin^2\theta-\Omega_2^2\cos^2\theta)^{1/2}}\right),\nonumber\\
    &\partial_{\theta}^2u \sim O\left(\frac{1}{(1-\Omega_1^2\sin^2\theta-\Omega_2^2\cos^2\theta)^{1/2}}\right)
\end{align}
are subleading to appear in the equation of motion.
The stress-energy tensor and charge density at the leading order are given as the same as that of \eqref{st}:
\begin{equation}
\begin{aligned}
    T_{tt}&=\frac{\mu^4(3+\Omega_1^2\sin^2\theta+\Omega_2^2\cos^2\theta)}{2\lambda(1-\Omega_1^2\sin^2\theta-\Omega_2^2\cos^2\theta)^{3}}, \\
    T_{t\phi}&=-\frac{2\mu^4\Omega_1\sin^2\theta}{\lambda(1-\Omega_1^2\sin^2\theta-\Omega_2^2\cos^2\theta)^{3}},\\
    T_{t\psi}&=-\frac{2\mu^4\Omega_2\cos^2\theta}{\lambda(1-\Omega_1^2\sin^2\theta-\Omega_2^2\cos^2\theta)^{3}},\\
    \rho&=j^t=\frac{2\mu^3}{\lambda(1-\Omega_1^2\sin^2\theta-\Omega_2^2\cos^2\theta)^{2}}.
\end{aligned}
\end{equation}
Also, note that $\langle\varphi\varphi\rangle\approx \frac{\mu^2}{\lambda(1-\Omega_1^2\sin^2\theta-\Omega^2\cos^2\theta)}$. This is consistent with the result obtained via the mean-field approximation in \eqref{C} for $\Omega_1=\Omega_2=\Omega$.

\section{Derivation of \eqref{deriv1}}\label{der}
Below is a self‑contained derivation of \eqref{deriv1}. On $S^3$, we are restricting to the highest‑weight sector
$$
m_1,m_2\;\ge0,
\quad
m_1 + m_2 = n,
$$
where the corresponding scalar harmonics take the simple form
\begin{align}
Y_{n,m_1,m_2}(\theta,\phi,\psi)
= N_{n,m_1}\,
e^{\,i\,(m_1\phi + m_2\psi)}\,
(\sin\theta)^{m_1}\,(\cos\theta)^{m_2},
\end{align}
with $\theta\in[0,\tfrac\pi2],~
\phi,\psi\in[0,2\pi),$
and $N_{n,m_1}$ chosen so that
$\displaystyle\int_{S^3}|Y|^2=1$.

The volume element on $S^3$ in these coordinates is
\begin{align}
d\Omega_3
=\sin\theta\,\cos\theta\;d\theta\,d\phi\,d\psi.
\end{align}
For fixed $(\phi,\psi)$, the $\theta$-dependence of $|Y_{n,m_1,m_2}|^2$ is
\begin{align}
|Y|^2
=|N_{n,m_1}|^2\,
(\sin\theta)^{2m_1}\,(\cos\theta)^{2m_2}.
\end{align}
Requiring $\int_{S^3}|Y|^2=1$ enforces
\begin{align}
\sum_{m_1=0}^n |N_{n,m_1}|^2
\int_0^{\frac\pi2}
(\sin\theta)^{2m_1+1}
(\cos\theta)^{2m_2+1}\,d\theta
=1.
\end{align}
But the integral is a Beta‐function, and one finds
\begin{align}
|N_{n,m_1}|^2
=\binom{n}{m_1}
\end{align}

(up to the common factor $1/(2\pi^2)$ which drops out in normalized sums).  In other words, at each fixed point $(\theta,\phi,\psi)$ the squared harmonics form a binomial distribution in $m_1$:
\begin{align}
\sum_{m_1=0}^n |Y_{n,m_1,n-m_1}|^2
=\sum_{m_1=0}^n \binom{n}{m_1}\,
(\sin\theta)^{2m_1}\,(\cos\theta)^{2(n-m_1)}
=1.
\end{align}

Since $\{\,|Y|^2\}_{m_1}$ is exactly the probability mass function of a $\mathrm{Binomial}(n,p)$ with
\begin{align}
p = \sin^2\theta,
\end{align}
we immediately have the standard moments:
\begin{align}
\sum_{m_1=0}^n m_1\,|Y|^2
= n\,p
= n\,\sin^2\theta,
\end{align}
and, because $m_2 = n-m_1$,
\begin{align}
\sum_{m_1=0}^n m_2\,|Y|^2
= \sum_{m_1=0}^n (n-m_1)\,|Y|^2
= n - n\,\sin^2\theta
= n\,\cos^2\theta.
\end{align}

\bibliography{References.bib}{}

\end{document}